\documentstyle[aps,twocolumn,epsfig]{revtex}

\begin{document}
\draft

\twocolumn[\hsize\textwidth\columnwidth\hsize\csname@twocolumnfalse\endcsname

\title{Violation of Bell inequality for thermal states of interaction qubits via a multi-qubit Heisenberg model}
\author{Xiaoguang Wang}
\address{Quantum information group, Institute for Scientific Interchange (ISI) Foundation,\\
 Viale Settimio Severo 65, 
I-10133 Torino, Italy}
\date{\today}
\maketitle

\begin{abstract}
We study the violations of Bell inequality for thermal states of qubits in a multi-qubit Heisenberg model as a function of temperature
and external magnetic fields. Unlike the behaviors of the entanglement the violation can not be obtained by increasing the temperature or the magnetic field. The threshold temperatures of the violation are found be less than that of the entanglement. We also consider a realistic cavity-QED model which is a 
special case of the mutli-qubit Heisenberg model.
\end{abstract}

]

Historically the violation of Bell inequalities \cite{Bell} was considered as
a means of determining whether there is entanglement between two qubits. In
1989 Werner \cite{Werner89} demonstrated that that there exists states which
are entangled but do not violate any Bell--type inequality, i.e., not all
entangled states violate a Bell inequality. Further studies \cite
{Popescu,Gisin} showed that the maximal violation of a Bell inequality does
not behave monotonously under local operation and classical communications.
So Bell violations only give a hint for entanglement and can be considered an
entanglement witness \cite{Terhal}. Most recently Scarani and Gisin built the
relations between Bell inequalities and the usefulness for quantum key
distribution \cite{QKD} and quantum secret sharing \cite{QSS}, and D\"{u}r
showed that even multipartite bound entangled states can violate Bell
inequalities \cite{Dur}.

An interesting and natural type of entanglement, thermal
entanglement, was introduced in magnetic systems 
and studied within the Heisenberg isotropic \cite
{Arnesen}, anisotropic Heisenberg models \cite{Wang,WangKlaus} and the Ising
model in a magnetic field \cite{Ising}. Another natural type of entanglement
was recently introduced in Bose-Einstein condensates \cite{Simon}.
The Gibbs state of a system at thermodynamic equilibrium is represented by the density operator 
\begin{equation}
\rho (T)=\exp \left( -H/kT\right) /Z,
\end{equation}
where $Z=$tr$\left[ \exp \left( -H/kT\right) \right] $ is the partition
function, $H$ the system Hamiltonian, $k$ is Boltzmann's constant which we
henceforth take equal to 1, and $T=1/\beta$ the temperature. As $\rho (T)$
represents a thermal state, the entanglement in the state is called {\em 
thermal entanglement}\cite{Arnesen}. At $T=0$, $\rho (T)$ represents the ground state which is pure for the non-degenerate case and mixed for the degenerate case. The ground state can be entangled but the thermal state $\rho (T)$ at $T=\infty$ can not be entangled as $\rho(T)$ is a completely random mixture. The entanglement in the ground state of isotropic Heisenberg models was  studied in the literature \cite{OConnor,Meyer}.  A recent interesting work\cite{QPT} showed that the success of the density matrix renormalization group method in understanding quantum phase transitions\cite{QPT1} is due to the way it preserves quantum entanglement under renormalization.

A complication in the analysis of thermal entanglement is that, although
standard statistical physics is characterized by the partition function,
determined by only the eigenvalues of the system, thermal entanglement
properties require in addition knowledge of the eigenstates. In other words
thermal entanglement is determined by both eigenstates and eigenvalues of
the density operator $\rho (T)$. An interesting behavior of the thermal
entanglement is that it can increase with the increase of temperature and
magnetic fields \cite{Arnesen,Wang}. A natural question is if we can induce the violation of Bell inequality by changing temperature and magnetic fields.

The study of thermal entanglement in the magnetic systems build a bridge between the quantum information theory and condensed matter physics. As there is intimate relations between the entanglement and the violation of Bell inequality, it will be interesting to investigate the Bell inequality in the magnetic systems, specifically in some well-known solvable quantum spin models such as Heisenberg models. 

Arnesen {\it et al.} \cite{Arnesen} pointed out that one can test the Clauser-Horne-Shimony-Holt (CHSH) inequality \cite{CHSH,Werner01} for the thermal state by measuring different components of the magnetic susceptibility tensor since the different components of the magnetic susceptibility tensor are proportional to spin-spin  correlations in different pairs of directions. 
Very recently Mancini and Bose \cite{Thermal} proposed a scheme for producing robust thermal entanglement between atoms in distant cavities connected by an optical fiber, and the scheme serves as an experiment to detect thermal entanglement when the cavities are very near.

In this paper we will study the violation of Bell inequality for thermal states of qubits which interact via a multi-qubit Heisenberg interaction model. One particular case of this model was recently realized in cavity-QED by Zheng \cite{ShiBiao}.
We will also discuss the violations in this realistic model.

The most commonly discussed Bell--inequality is the CHSH inequality.
The CHSH operator ($\vec{a},\vec{a^\prime},\vec{b},\vec{b^\prime}$ are unit vectors) reads

\begin{equation}
\hat{B}=\vec{a}\cdot \vec{\sigma}\otimes (\vec{b}+\vec{b^\prime})\cdot 
\vec{\sigma}+\vec{a^\prime}\cdot \vec{\sigma}\otimes (\vec{b}-\vec{b^\prime}%
)\cdot \vec{\sigma}.
\end{equation}
In the above notation, the Bell inequality reads

\begin{equation}
\left| \langle \hat{B}\rangle \right| =\left| \text{tr}(\rho \hat{B})\right|
\leq 2,
\end{equation}
where $\rho $ is an arbitrary two--qubit state. The maximal amount of Bell
violation of a state $\rho $ is given by\cite{Horo}

\begin{equation}
{\cal B}=2\sqrt{u+\tilde{u}},
\end{equation}
where $u$ and $\tilde{u}$  are the two largest eigenvalues of 
$T_\rho T_\rho ^{\dagger },$ $T_\rho $ being the $3\times 3$ matrix whose
elements is 
\begin{equation}
(T_\rho)_{nm}=\text{tr}(\rho \sigma _n\otimes \sigma _m).
\label{eq:ele}
\end{equation}
Here $\sigma_{1}\equiv \sigma_{x}, \sigma_{2}\equiv\sigma_{y}, $ and $\sigma_{3}\equiv\sigma_{z}$ are the usual Pauli matrices.

Now we define a
quantity 
\begin{equation}
M=u+\tilde{u}-1  \label{eq:m}
\end{equation}
which ranges from -1 to 1. The quantity $M,\ $which we call {\em violation
measure} of Bell inequality in this paper, indicates the Bell violation when $M>0$ and maximal Bell violation when $M=1$.

For any pure and some special mixed two-qubit state, there exists a simple relation \cite{Munro} between the concurrence \cite{Conc} $C$ (the measure of entanglement) and the maximal violation of the Bell inequality:
\begin{equation}
{\cal B}=2\sqrt{1+C^2}.
\end{equation} 
We see from the above equation that the entanglement implies the violation of Bell inequality. But for a general mixed state, there is no this simple relation.

We consider a $N$--qubit Heisenberg model
\begin{eqnarray}
H &=&\frac J4\sum_{i\neq j}^N\left( \sigma _{ix}\otimes \sigma _{jx}+\sigma
_{iy}\otimes \sigma _{jy}+\Delta \sigma _{iz}\otimes \sigma _{jz}\right) 
\nonumber \\
&&+\frac B2\sum_{i=1}^N\sigma _{iz},\label{eq:h}
\end{eqnarray}
where $J$ is the exchange coupling contant, $\Delta $ is the anisotropic
parameter, and $B$ is the magnetic field along $z$ direction.
In this model all particles interact equally with each other. 
The model reduces to  the two-qubit model for $N=2$ and to the three--qubit Heisenberg model with the periodic boundary condition for $N=3$, respectively.

By using the collective spin operators $S_\alpha =\sum_{i=1}^N\sigma
_{i\alpha }/2$ $(\alpha =x,y,z)$, the Hamiltonian is rewritten as 
\begin{eqnarray}
H &=&J\left( S_x^2+S_y^2+\Delta S_z^2\right) +BS_z  \nonumber \\
&=&J\vec{S}^2+J(\Delta -1)S_z^2+BS_z  \label{eq:hh}
\end{eqnarray}
up to a trivial constant. Due to the key term $S_z^2$ one can generate multipartite maximally entangled states in this system \cite{Klaus,ShiBiao}

The partition function is obtained as \cite{WangKlaus} 
\begin{eqnarray}
Z &=&\sum_{k=0}^{N/2}N_k\sum_{m=0}^{N-2k}e^{-\beta J[(\Delta
-1)(m-N/2+k)^2]}e^{-\beta B(m-N/2+k)}  \nonumber \\
&&\times e^{-\beta J(N/2-k)(N/2-k+1)},  \label{eq:zz}
\end{eqnarray}
where $N_k=\left( 
\begin{array}{c}
N \\ 
k
\end{array}
\right) -\left( 
\begin{array}{c}
N \\ 
k-1
\end{array}
\right) $ and $\left( 
\begin{array}{c}
N \\ 
-1
\end{array}
\right) =0.$

The reduced density matrix for any two qubits is given by \cite{WangKlaus}

\begin{eqnarray}
\rho _{12} &=& v_+|00\rangle\langle 00| +v_-|11\rangle\langle 11 |
+ w(|01\rangle\langle 01| +|10\rangle\langle 10 |)\nonumber\\
&& +y(|01\rangle\langle 10| +|10\rangle\langle 01 |)\nonumber\\
&=&\frac 14[1+(v_{+}-v_{-})(\sigma _{1z}+\sigma _{2z})  \nonumber
\\
&&+2y(\sigma _{1x}\sigma _{2x}+\sigma _{1y}\sigma _{2y})+(1-4w)\sigma
_{1z}\sigma _{2z}]  \label{eq:rhorho}
\end{eqnarray}
with matrix elements

\begin{eqnarray}
v_{\pm } &=&\frac{N^2-2N+4\langle S_z^2\rangle \pm 4\langle S_z\rangle (N-1)%
}{4N(N-1)},  \nonumber  \label{eq:aaa} \\
w &=&\frac{N^2-4\langle S_z^2\rangle }{4N(N-1)},  \nonumber \\
y &=&\frac{2\langle S_x^2+S_y^2\rangle -N}{2N(N-1)}.  \label{eq:bbb}
\end{eqnarray}

The relevant expectation values are given by\cite{WangKlaus}

\begin{eqnarray}
\langle f(S_z)\rangle  &=&\sum_{k=0}^{N/2}N_k\sum_{m=0}^{N-2k}f(m-N/2+k) 
\nonumber \\
&&\times e^{-\beta J[(\Delta -1)(m-N/2+k)^2]}e^{-\beta B(m-N/2+k)}  \nonumber
\\
&&\times e^{-\beta J(N/2-k)(N/2-k+1)}/Z,  \nonumber \\
\langle S_x^2+S_y^2\rangle 
&=&\sum_{k=0}^{N/2}N_k\sum_{m=0}^{N-2k}[(N/2-k)(N/2-k+1)  \nonumber \\
&&-(m-N/2+k)^2]  \nonumber \\
&&\times e^{-\beta J[(\Delta -1)(m-N/2+k)^2]}e^{-\beta B(m-N/2+k)}\text{ } 
\nonumber \\
&&\times e^{-\beta J(N/2-k)(N/2-k+1)}/Z.  \label{eq:xyz}
\end{eqnarray}
Here $f(S_z)$ is an arbitrary function of $S_z$. Eqs.(\ref{eq:zz})--(\ref{eq:xyz}) gives completely the reduced density matrix. Along the standard procedures for calculating the maximal violation and the concurrence, they are obtained as
\begin{eqnarray}
M &\equiv&M(J,\Delta,\beta,B)\nonumber\\
&&=8y^2+(1-4w)^2-\min [4y^2,(1-4w)^2]-1  \label{eq:m2} \\
C &\equiv&C(J,\Delta,\beta,B)=2\max \{0,|y|-\sqrt{v_{+}v_{-}}\}  \label{eq:c2}
\end{eqnarray}

For $N=2$, Eqs.(\ref{eq:m2}) and (\ref{eq:c2}) become
\begin{eqnarray}
M &=&\{2\sinh ^2(\beta J)+[\cosh (\beta B)e^{-\beta \Delta J}-\cosh (\beta
J)]^2  \nonumber \\
&&-\min \{\sinh ^2(\beta J),[\cosh (\beta B)e^{-\beta \Delta J}-\cosh (\beta
J)]^2\}\}  \nonumber \\
&&/[\cosh (\beta B)e^{-\beta \Delta J}+\cosh (\beta J)]^2-1, \\
C &=&\max \{0,\sinh (\beta |J|)-e^{-\beta \Delta J}\}  \nonumber \\
&&/[\cosh (\beta B)e^{-\beta \Delta J}+\cosh (\beta J)].
\end{eqnarray}

From the analytical expressions of $M$
and $C$ we immediately obtain
\begin{eqnarray}
M(J,\Delta,\beta,B)&=& M(-J,-\Delta,\beta,B), \nonumber\\
C(J,\Delta,\beta,B)&=& C(-J,-\Delta,\beta,B), \label{eq:mc1}\\ 
M(J,\Delta,\beta,B)&=& M(J,\Delta,\beta,-B), \nonumber\\
C(J,\Delta,\beta,B)&=& C(J,\Delta,\beta,-B). \label{eq:mc2}
\end{eqnarray}
From Eq.(\ref{eq:mc1}) we see that the violation of Bell inequality is invariant
under the simultaneous sign changes of the parameters $\Delta $ and $J$. The entanglement also has the same property. In
particular the violation does not depend on the sign of $J$ when $\Delta =0$ which implies that the violations are same for the antiferromagnetic ($J>0$) and
ferromagnetic cases ($J<0$) in the $XX$ model. 
From Eq.(\ref{eq:mc2}) we know that $M$ are symmetric with respect to $B$. Although this conclusion is obtained for the case of $N=2$, it is valid for any $N$ in our model.

Fig.1 gives numerical calculations of the violation measure and concurrence as a function of temperature for different magnetic fields. The concurrence is plotted in this paper for a comparison. It is clear from Fig.1 that there exist threshold temperatures $T_M$ ($T_C$) for the violation measure (the entanglement). After the threshold temperature $T_M$ ($T_C$)  the Bell inequality is not violated (the entanglement disappears). Note that $T_C$ is independent on the magnetic fields, while $T_M$ decreases with the increase of magnetic fields. We also observe that
$T_M < T_C$ which implies the fact that some entangled states do not violate a Bell inequality. In the temperature range $T_M<T<T_C$, the state is entangled but the Bell inequality is not violated.
Now we see the case of $B=2.5$. The entanglement can be increased by increasing the temperature, however the Bell inequality is not violated at any temperatures. The comparison shows that the behavior of the violation measure is very different from that of the concurrence.

\begin{figure}
\begin{center}
\epsfig{width=10cm,file=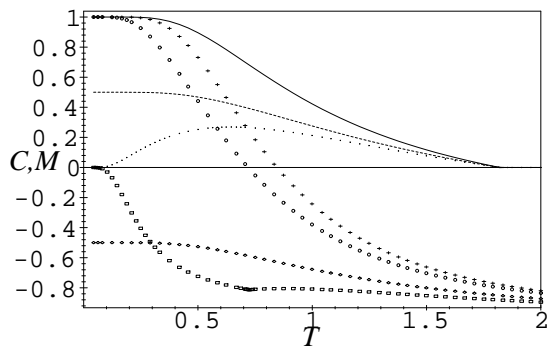}
\caption[]{The violation measure and the concurrence as a function of 
temperature for different magnetic fields. Violation measure: $B=0$ (cross points), $B=1$ (circle points), $B=2$ (diamond points) and $B=2.5$ (box points);  
The concurrence: $B=0$ (solid line)
, $B=2$ (dashed line), and $B=2.5$ (dotted line). 
The parameters $J=\Delta=1$ and $N=2$.} 
\end{center}
\end{figure}

In Fig.2 we plot the violation measure as function of temperature for different number of qubits. The results show that although we can have entanglement the Bell inequality can not be violated for $N\ge 3$ as seen clearly from the figure that $M<0$. When $N\ge 3$ we obtain the two-qubit reduced density matrix by tracing out all the other qubits but two of them. This usually increases the entropy and make the Bell inequality much less be violated. We also see that the threshold temperature $T_C$ increases as the number of qubits $N$ increases and $T_M<T_C$.

Then we see if the magnetic fields can induce the violation for $N\ge 3$. The violation measure versus the magnetic field is given in Fig. 3. The temperature is choose to be close to the absolute zero temperature. From Fig.3 we see that we still can not obtain the violation of Bell inequality for $N\ge 3$ by increasing the magnetic fields although the entanglement can be increased by the magnetic fields. For the case of $N=2$ we find the threshold magnetic field $B_M$ for the violation measure and $B_C$ for the concurrence. The threshold magnetic field $B_M$ is less than $B_C$. We also find that the threshold magnetic field increases as the number of qubits increases. The magnetic field can increase the entanglement. The more complicated behaviors of the entanglement can be found for the case of 6 qubits (dotted line). There are two dips when the magnetic field increases. After the two dips, the entanglement increases and then decreases and reach the threshold point.

\begin{figure}
\begin{center}
\epsfig{width=10cm,file=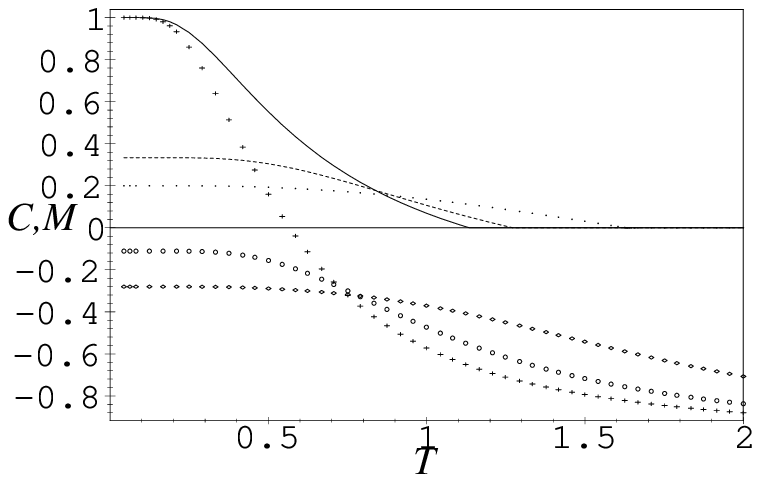}
\caption[]{The violation measure and the concurrence as a function of 
temperature for different number of qubits. The violation measure: $N=2$ (cross points), $N=3$ (circle points), and $N=5$ (diamond points); The concurrence: $N=2$ (solid line)
, $N=3$ (dashed line), and $N=5$ (dotted line). 
The parameters $J=-1$ and $B=\Delta=0$.} 
\end{center}
\end{figure}

\begin{figure}
\begin{center}
\epsfig{width=10cm,file=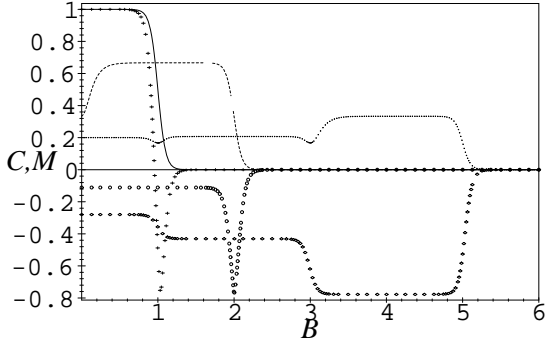}
\caption[]{The violation measure and the concurrence as a function of 
the magnetic field for different number of qubits. The violation measure: $N=2$ (cross points), $N=3$ (circle points), and $N=6$ (diamond points); The concurrence: $N=2$ (solid line)
, $N=3$ (dashed line), and $N=6$ (dotted line). 
The parameters $J=-1$, $T=0.05$, and $\Delta=0$.} 
\end{center}
\end{figure}

Another interesting feature of quantum entanglement is that a bipartite entangled state (systems $A$ and $B$) can be more disordered locally than globally \cite{Nielsen01}. If we measure the disorder by von Neumann entropy $S(\rho)=-\text{tr}(\rho \log_2 \rho)$, one inequality holds for all separable states
\begin{equation}
S(A), S(B) \le S(A,B). \label{eq:sss}
\end{equation}
which is called disorder inequality in this paper.
For the models we are considering, $S(A)=S(B)$ holds. For the general case $S(A)$ is usually not equal to $S(B)$. 
Then we define a quantity
\begin{equation}
D=S(A)-S(A,B),
\end{equation}
which can be larger than zero and acts as a quantitative measure of the violation of the disorder inequality (\ref{eq:sss}) Interestingly the quantity $D$ gives directly a lower bound of the entanglement of formation $E_f(A,B)$ of the state\cite{Horodecki},
i.e., 
\begin{equation}
E_f(A,B)\ge D,
\end{equation}
which shows that the state is entangled if $D>0$.

From Eq.(\ref{eq:rhorho}), the violation measure $D$ is obtained as
\begin{eqnarray}
D&=&v_+\log_2 v_+ + v_-\log_2 v_-+(w-y)\log_2 (w-y)\nonumber\\
&&+ (w+y)\log_2 (w+y)
-(v_++w)\log_2 (v_++w)\nonumber\\
&&- (v_-+w)\log_2 (v_-+w).
\end{eqnarray}

\begin{figure}
\begin{center}
\epsfig{width=10cm,file=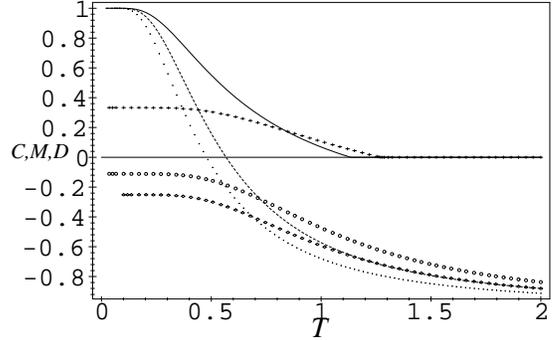}
\caption[]{The violation measure $D$ of the disorder inequality , the concurrence $C$ and violation measure $M$ as a function of $T$.
We plotted $C$ (solid line), $M$ (dashed line), and $R$ (dotted line)
for $N=2$ and $C$ (cross points), $M$ (circle points), and $R$ (diamond points) for $N=3$, respectively. 
The parameters $J=-1$ and $B=\Delta=0$.} 
\end{center}
\end{figure}
The numerical results are given in Fig.4. We find that the behaviors of the violation measures are similar. For $N=2$, both the disorder inequality and Bell inequality can be violated. But the threshold temperature $T_D$ of the violation measure is different from $T_M$ and $T_C$, i.e., $T_D<T_M<T_C$.
There exists a temperature range $T_D<T<T_M$ in which the Bell inequality is violated while the disorder inequality is not violated. For $N=3$ we still can not find any violation of both the disorder and Bell inequality although the state is entangled for a range of temperature. 

Finally we consider a realistic model in cavity QED. Consider the interaction of 
$N$ qubits with a single cavity mode in the vacuum state via Tavis-Cumming model \cite{TV}.  For the case of large detuning $\delta$ between qubits and the field mode, the effective Hamiltonian is given by \cite{ShiBiao}
\begin{equation}
H_e=\lambda(S_x^2+S_y^2+S_z), \label{eq:qed}
\end{equation}
where $\lambda={g^2}/{\delta}$ and $g$ is the atom-cavity coupling strength. Comparing Eqs.(\ref{eq:qed}) and (\ref{eq:h}) we find $\Delta=0$ and $J=B=\lambda$. Note that $\lambda$ can be negative due to the detuning. The numerical results are shown in Fig.5. We see that the entanglement can be obtained while the Bell and disorder inequality can not be violated. The 
behavior of this physical model is similar to that obtained from the `abstract'
model. Both the analytical and numerical results show that the Bell inequality is more difficult to be obtained than the entanglment.

\begin{figure}
\begin{center}
\epsfig{width=10cm,file=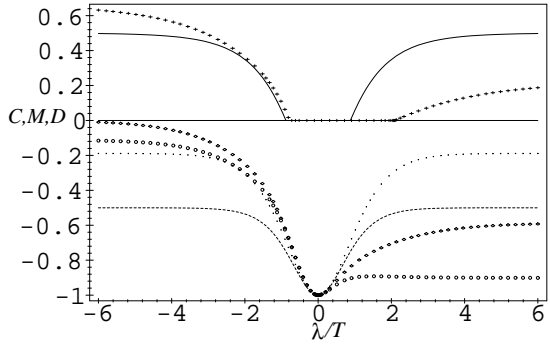}
\caption[]{The violation measures and the concurrence as a function of $\lambda/T$.
We plotted $C$ (solid line), $M$ (dashed line), and $R$ (dotted line)
for $N=2$ and $C$ (cross points), $M$ (circle points), and $R$ (diamond points) for $N=3$, respectively.} 
\end{center}
\end{figure}

We consider the multi-qubit thermal state in the multi-qubit Heisenberg model which is exactly solvable. We trace out all the other qubits but two. Then we obtain the two-qubit reduced density matrix, from which we have studied the violation of Bell inequality. The results show that the Bell inequality are violated for the case of $N=2$ but not violated for the case of $N\ge 3$ at any temperature and magnetic fields in our model. Unlike the behaviors of the entanglement, the violation can not be induced by increasing temperature and magnetic fields. In comparison with the thermal entanglement, they  are relatively hard to be obtained. 

\acknowledgments
The author thanks Klaus M\o lmer, Hongchen Fu, Allan I Solomon, Ujjwal Sen, Mang Feng, R. de la Madrid, Barry C Sanders and Paolo Zanardi 
for helpful discussions. I thank Prof. Guenter Mahler and Irene D'Amico for useful comments. This work has been supported by the European
Community through grant ISI-1999-10596 (Q-ACTA).

\end{document}